\documentclass[12pt]{article}
\usepackage{graphicx}

\def\napoli{Institute of Nuclear Physics Polish Academy of Sciences, Radzikowskiego 152 st., Cracow, Poland}
\def\support{\footnote{This was partially supported by Polish Ministry of Science and Higher Education under the Mobility Plus programme (1285/MOB/IV/2015/0) and by Polish National Science Centre under the grant number UMO-2015/17/D/ST2/03530.}}

\def\Title#1{\begin{center} {\Large #1 } \end{center}}
\def\Author#1{\begin{center}{ \sc #1} \end{center}}
\def\Address#1{\begin{center}{ \it #1} \end{center}}

\newenvironment{Abstract}{\begin{quotation}  }{\end{quotation}}
\newenvironment{Presented}{\begin{quotation} \begin{center} 
             PRESENTED AT\end{center}\bigskip 
      \begin{center}\begin{large}}{\end{large}\end{center} \end{quotation}}

\def\beq{\begin{equation}}
\def\eeq#1{\label{#1}\end{equation}}
\def\eeqn{\end{equation}}

\def\beqa{\begin{eqnarray}}
\def\eeqa#1{\label{#1}\end{eqnarray}}
\def\eeqan{\end{eqnarray}}

\let\bar=\overbar

\def\Dslash{\not{\hbox{\kern-4pt $D$}}}
\def\dslash{\not{\hbox{\kern-2pt $\del$}}}

\def\msb{{\bar{\ssstyle M \kern -1pt S}}}

\begin{document}
\begin{titlepage}

\vfill
\Title{GenEx -- Exclusive Meson Generator}
\vfill
\Author{Maciej Trzebi\'nski\support}
\Address{\napoli}
\vfill
\begin{Abstract}
Exclusive light meson production processes possible to be measured at RHIC and LHC accelerators are briefly described. This includes Pomeron and photon induced continuum of pions and kaons as well as $f_0$ and $\rho^0$ resonant production. Next, GenEx, a new Monte Carlo generator recently developed in Cracow, is presented. Its purpose is to generate resonant and non-resonant pion and kaon production and diffractive Bremsstrahlung. Its future development will take into account spin (polarization) effects and simulate corrections due to absorption and re-scattering.
\end{Abstract}
\vfill
\begin{Presented}
Presented at EDS Blois 2017, Prague, \\ Czech Republic, June 26-30, 2017
\end{Presented}
\vfill
\end{titlepage}
\def\thefootnote{\fnsymbol{footnote}}
\setcounter{footnote}{0}

\section{Introduction}

Collisions at hadron accelerators are dominated by soft processes. Absence of a hard scale in these events prevents one from using perturbation theory. Instead, in order to calculate the properties of the produced particles (\textit{e.g.} energy or angular distributions), one has to use approximative methods. 

In about a half of collisions at LHC or RHIC one or both outgoing protons stay intact. Such final state is possible due to the exchange of a colourless object: photon (electromagnetic) or Pomeron (strong interaction). Such an exchange results in a presence of the rapidity gap between the centrally produced system and scattered protons. Thus, such events are of diffractive nature. 

Diffractive studies are one of the important parts of the physics programme for the LHC experiments. A particularly interesting classes are exclusive processes, where all centrally-produced particles are detected. An example is the process of pion pair production $pp \rightarrow p \pi^+ \pi^- p$, which can be used for studies of low-mass resonances, including searches for glueballs.

In order to include detector effects (acceptance, efficiency) in theory-to-data comparison Monte Carlo generators are needed. There are few MC generators which are specialized in simulation of the soft exclusive processes, \textit{e.g.} \textsc{SuperCHIC}~\cite{SuperCHIC} or \textsc{DIME}~\cite{DIME}. In this paper \textsc{GenEx}~\cite{GenEx1, GenEx2}, a tool complementary to the existing ones in terms of implemented processes and calculation methods, will be briefly described.

\section{Soft Exclusive Production at the LHC}

Diffractive protons are usually scattered at very small angles (fraction of milliradian). In order to measure them, special devices that allow detector presence in close vicinity of the beam are required. At the LHC Roman pots are commonly used. ATLAS~\cite{ATLAS} has two systems of such detectors: ALFA~\cite{ALFA1, ALFA2} and AFP~\cite{AFP}. Similar set is installed around CMS/TOTEM Interaction Point~\cite{TOTEM}. Roman pots are usually located far away (few hundreds meters) from the Interaction Point (IP). It means that there are several LHC magnets between them and the IP. The settings of these magnets, commonly called \textit{machine optics}, play a key role in the exclusive analysis. The detailed description of the properties of optics sets used at the LHC can be found in~\cite{LHC_optics}. In addition, in this reference a geometric acceptance of AFP and ALFA detectors is presented.

\subsection{Non-resonant Pion Pair Production}
\begin{figure}[htb]
  \centering
  \includegraphics[width=0.25\textwidth]{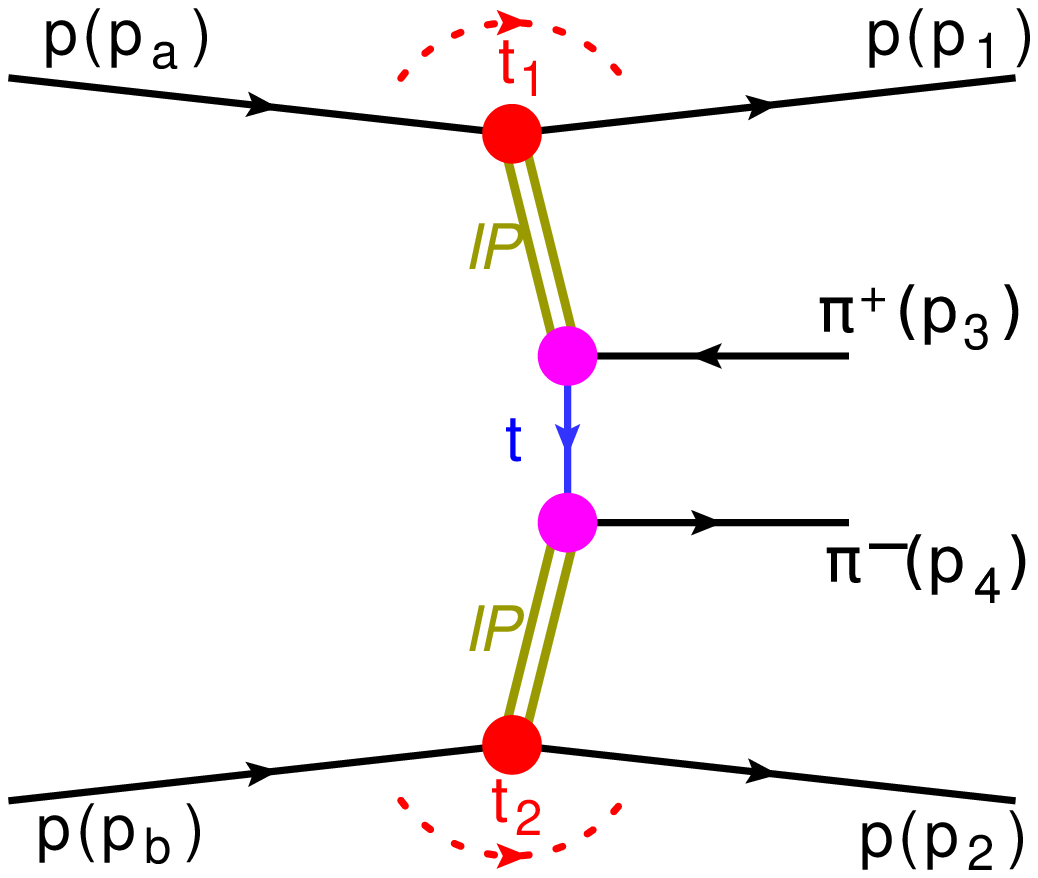}\hspace{3cm}
  \includegraphics[width=0.25\textwidth]{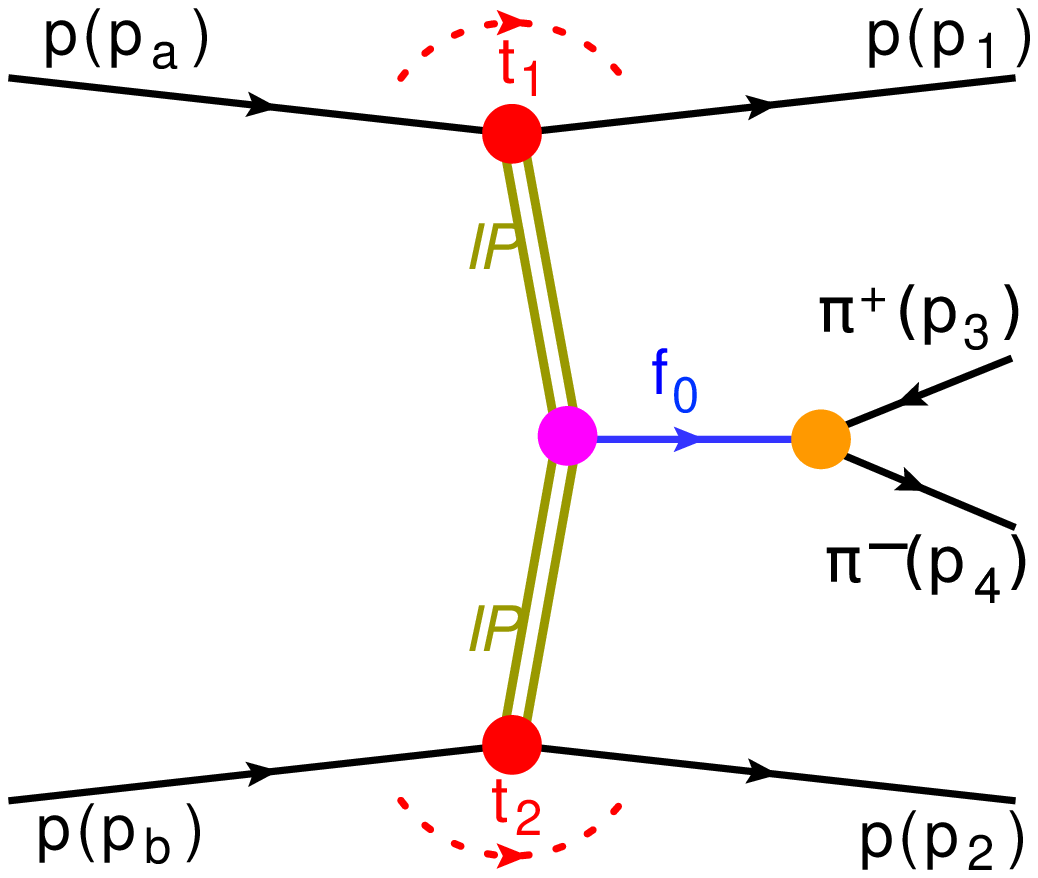}
  \caption{Diagrams of: non-resonant (\textbf{left}) and $f_0$ resonance (\textbf{right}) pion pair production.}
  \label{fig1}
\end{figure}

Diagram for the non-resonant pion pair production (also called continuum) is shown in Fig.~\ref{fig1} (left); a Pomeron is ``emitted'' from each proton resulting in a four particles present in final state: scattered protons and (central) pions. The object exchanged in the $t$-channel is an off-shell pion. The model implemented in GenEx is based on calculations done in Ref.~\cite{Lebiedowicz1}. Feasibility studies for the ATLAS detector are shown in~\cite{exc_pions}.

\subsection{Pion Pair Production via $f_0$ Resonance}
Diagram for the exclusive pion pair production via the $f_0$ resonance is shown in Fig.~\ref{fig1} (right). In order to calculate it, one must know the Pomeron-Pomeron-meson vertices. These elements have been discussed in detail in Ref.~\cite{Lebiedowicz2}. It was shown there that different couplings structures with corresponding orbital angular momentum and spin of two Pomerons are possible.

\subsection{Photon-induced Continuum and $\rho^0$ Photoproduction}
\begin{figure}[htb]
  \centering
  \includegraphics[width=0.25\textwidth]{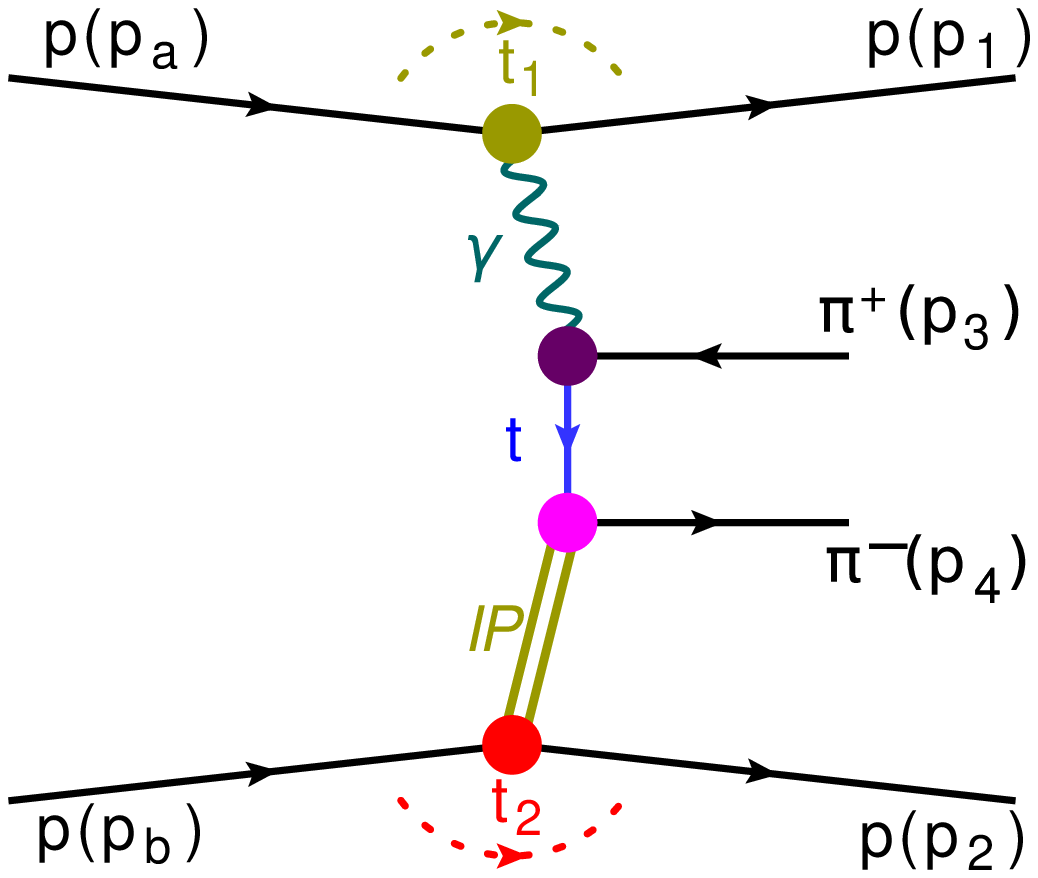}\hspace{3cm}
  \includegraphics[width=0.25\textwidth]{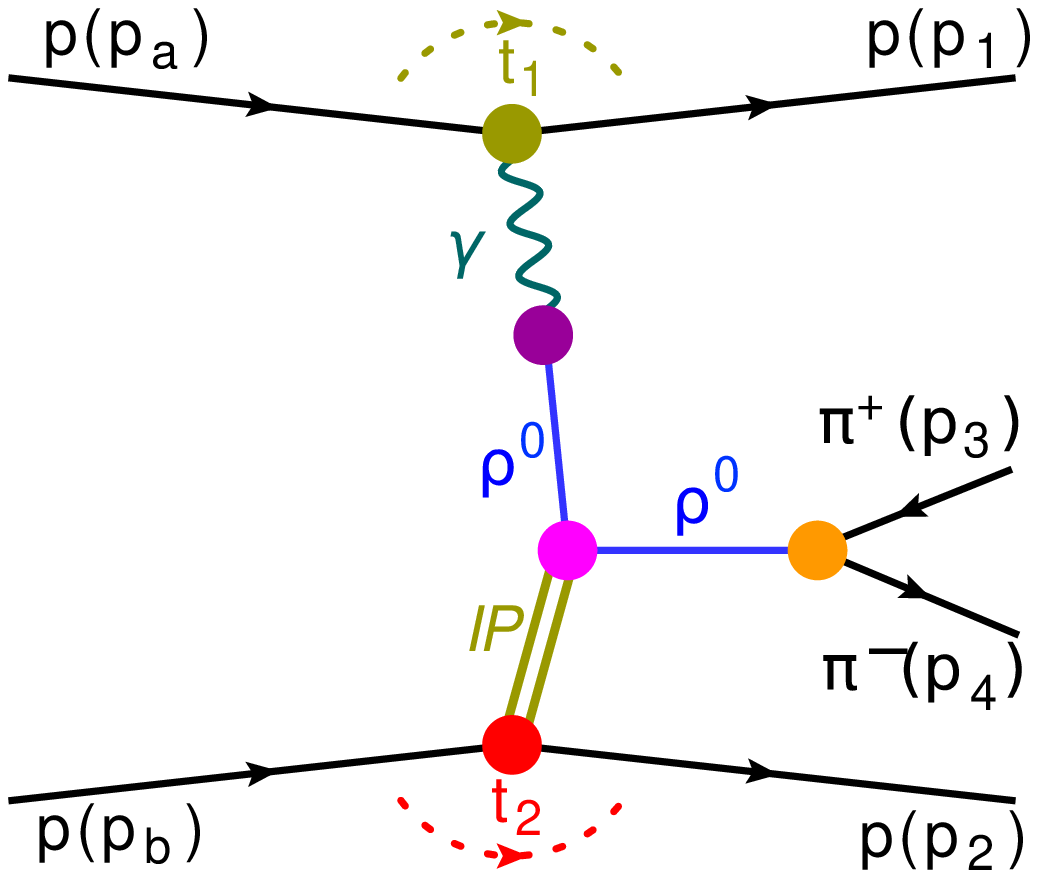}
  \caption{Diagrams of: non-resonant (\textbf{left}) and $\rho^0$ (\textbf{right}) pion pair photo-production.}
  \label{fig2}
\end{figure}

The dominant diagram of the exclusive pion pair continuum production is a Pomeron-induced one. However, the production of a photon-induced continuum (see Fig.~\ref{fig2} (left)) is also possible. It should be taken into account in the Monte Carlo generators as it contributes up to the several percent of the exclusive pion pair production, see Ref.~\cite{exc_pions_photon}.

Another interesting class of photon-induced processes is a resonant $\rho^0$ photoproduction. Feynman diagram is shown Fig.~\ref{fig2} (right) whereas details are discussed in ~\cite{exc_pions_photon}.

\subsection{Diffractive Bremsstrahlung}

\begin{figure}[htb]
  \centering
  \includegraphics[width=0.35\textwidth]{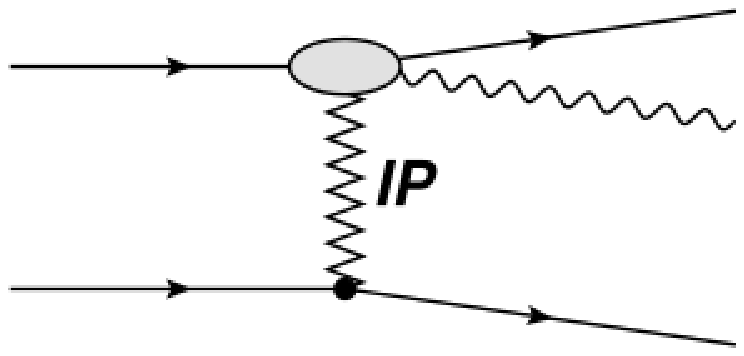}\hspace{2cm}
  \includegraphics[width=0.35\textwidth]{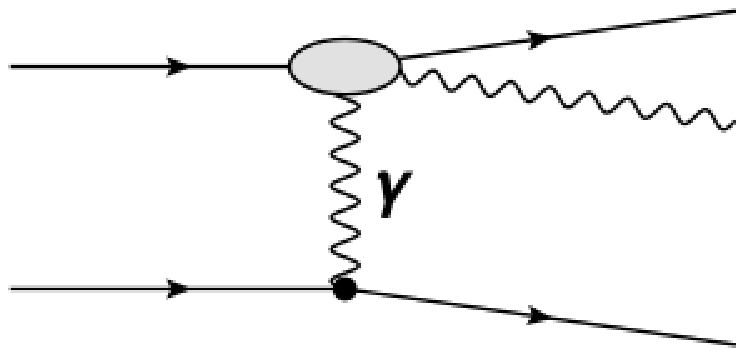}
  \caption{Diagrams of electromagnetic (left) and diffractive (right) Bremsstrahlung. The blobs represent various mechanisms of photon emission.}
  \label{fig3}
\end{figure}

Diffractive Bremsstrahlung is typically an electromagnetic process (see Fig.~\ref{fig3} (left)). However, as discussed in~\cite{brem1}, high energy photons can be radiated in the elastic proton-proton scattering (Fig.~\ref{fig3} (right)). This idea was extended in~\cite{brem2} by introducing the proton form-factor into the calculations and by considering also other mechanisms leading to the $pp\gamma$ final state, \textit{e.g.} a virtual photon re-scattering. Recently, diffractive Bremsstrahlung was implemented into the GenEx Monte Carlo generator~\cite{brem3}.

\section{GenEx Generator}

\textsc{GenEx}~\cite{GenEx1} is a C++ class structure for the construction of a Monte Carlo event generators which can produce unweighted events within the relativistic phase space. Generator is self-adapting to the provided matrix element and acceptance cuts. Current version contain resonant and non-resonant exclusive meson production processes with vector and tensor Pomeron as well as diffractive Bremsstrahlung production~\cite{GenEx3}. In near future spin (polarization) effects and absorption/re-scattering corrections will be added.

Current version of generator is based on several core classes:
\begin{itemize}
  \item \textsc{TFoam} -- class of adaptive Monte Carlo simulator,
  \item \textsc{TDensity} -- class with integrand function (calculated accordingly to a given event),
  \item \textsc{TEventMaker2toN} -- generates two leading particles and a central blob which then is decayed by \textsc{TDecay} into $N - 2$ remaining particles.
\end{itemize}
Recently, a significant effort is taken in order to simplify the implementation of new processes. For reference see~\cite{Ciesla}.

\section{Summary}
Exclusive light meson production should be measurable at LHC and RHIC. A Monte Carlo generator containing continuum and resonances would be useful to understand recorded data.  Such tool will be especially needed in order to include detector effects (acceptance, efficiency) in theory-data comparison.

One of such generators is \textsc{GenEx}. Its first version is already available for tests (see~\cite{GenEx3}) and contains: the generator structure, pion/kaon continuum and diffractive Bremsstrahlung processes. The developments are ongoing -- in the near future a new version of \textsc{GenEx} would allow to: generate resonant production ($f_0(500)$, $f_0(980)$, $f_0(1370)$, $f_0(1500)$, $f_2(1270)$, $f_2'(1520)$ and $\rho_0$), take into account spin (polarization) effects and simulate corrections due to absorption and re-scattering.


\begin{thebibliography}{99}

\bibitem{SuperCHIC} L. Harland-Lang \textit{et al.}, \textit{SuperCHIC: Monte Carlo Event Generator for Central Exclusive Production}, https://superchic.hepforge.org/.
%
\bibitem{DIME} L. Harland-Lang \textit{et al.}, \textit{Dime Monte Carlo: event generator for exclusive meson pair production via double Pomeron exchange}, http://dimemc.hepforge.org/.
%
\bibitem{GenEx1} R. A. Kycia, J. Chwastowski, R. Staszewski, J. Turnau, \textit{GenEx: A simple generator structure for exclusive processes in high energy collisions}, arXiv:1411.6035.
%
\bibitem{GenEx2} P. Erland, R. Staszewski, M. Trzebinski, R. Kycia, \textit{Elastic Hadron Scattering in Various Pomeron Models}, Acta Phys. Pol. B \textbf{48} (2017) 981.
%
\bibitem{ATLAS} ATLAS Collaboration, \textit{The ATLAS Experiment at the CERN Large Hadron Collider}, J. Instrum. \textbf{3} (2008) S08003.
%
\bibitem{ALFA1} ATLAS Luminosity and Forward Physics Community, \textit{Technical Design Report}, CERN-LHCC-2008-004, https://cds.cern.ch/record/1095847.
%
\bibitem{ALFA2} S. Abdel Khalek \textit{et al.}, \textit{The ALFA Roman Pot Detectors of ATLAS}, JINST \textbf{11} (2016) P11013.
%
\bibitem{AFP} ATLAS Collaboration, \textit{Letter of Intent for the Phase-I Upgrade of the ATLAS Experiment}, CERN-LHCC-2011-012, https://cds.cern.ch/record/1402470.
%
\bibitem{TOTEM} TOTEM Collaboration, \textit{Technical Design Report}, CERN-LHCC-2004-002,\\
TOTEM Collaboration, \textit{TOTEM Upgrade Proposal}, CERN-LHCC-2013-009.
%
\bibitem{LHC_optics} M. Trzebinski, \textit{Machine Optics Studies for the LHC Measurements}, in proceedings of XXXIV-th SPIE Joint Symposium Wilga vol. 9290 26. Proc. SPIE Int. Soc. Opt. Eng., 0277, 786; arXiv:1408.1836
%
\bibitem{Lebiedowicz1} P. Lebiedowicz, O. Nachtmann, A. Szczurek, \textit{Central exclusive diffractive production of $\pi^+\pi^-$ continuum, scalar and tensor resonances in $pp$ and $p\bar{p}$ scattering within tensor pomeron approach}, Phys. Rev. D \textbf{93} (2016) 054015.
%
\bibitem{exc_pions} R. Staszewski, P. Lebiedowicz, M. Trzebinski, J. Chwastowski, A. Szczurek, \textit{Exclusive $\pi^+\pi^-$ Production at the LHC with Forward Proton Tagging}, Acta Physica Polonica B \textbf{42} (2011) 1861.
%
\bibitem{Lebiedowicz2} P. Lebiedowicz, O. Nachtmann, A. Szczurek, \textit{Exclusive central diffractive production of scalar and pseudoscalar mesons tensorial vs. vectorial pomeron}, Annals Phys. \textbf{344} (2014) 301-339.
%
\bibitem{exc_pions_photon} P. Lebiedowicz, O. Nachtmann, A. Szczurek, \textit{The $\rho^0$ and Drell-Soding contributions to central exclusive production of $\pi^+\pi^-$ pairs in proton-proton collisions at high energies}, Phys. Rev. D \textbf{91} (2015) 074023.
%
\bibitem{brem1} V. A. Khoze, J. W. Lamsa, R. Orava, M. G. Ryskin, \textit{Forward Physics at the LHC: Detecting Elastic pp Scattering by Radiative Photons}, JINST \textbf{6} (2011) P01005.
%
\bibitem{brem2} P. Lebiedowicz and A. Szczurek, \textit{Exclusive diffractive photon bremsstrahlung at the LHC}, Phys. Rev. D \textbf{87} (2013) 114013.
%
\bibitem{brem3} J. J. Chwastowski, S. Czekierda, R. Staszewski, M. Trzebinski, \textit{Diffractive Bremsstrahlung at High-$\beta^*$ LHC Case Study}, Eur. Phys. J. C \textbf{77} (2017) 216.
%
\bibitem{GenEx3} \texttt{https://github.com/rkycia/GenEx}.
%
\bibitem{Ciesla} P.~Buglewicz, K.~Cie\'sla, \textit{Calculations in tensor pomeron model -- an object-oriented implementation}, talk at Challenges in Photon Induced Interactions, Cracow, 5-8 Sep. 2017.
\end{thebibliography}
\end{document}